# Dual effect of surface plasmons in light transmission through perforated metal films

Cheng-ping Huang[1,2], Qian-jin Wang[1], and Yong-yuan Zhu[1,*]

[1]*National Laboratory of Solid State Microstructures, Nanjing University*
*Nanjing 210093, P.R. China*
[2]*Department of Applied Physics, Nanjing University of Technology*
*Nanjing 210009, P.R. China*

The light transmission through square and rectangular holes in a metal film has been studied. By taking account of plasma response of real metal on hole walls as well as metal surface, an analytical result for the transmission has been deduced, which agrees well with the experiments. We show that the light transmission involves both the diffraction modes and the surface plasmons. Strong coupling of diffraction modes to surface plasmon polariton results in the transmission minima, whereas the coupling to cavity surface plasmon leads to the transmission maxima. The results suggest a dual-effect of the surface plasmons. [Phys.Rev. B **75**, 245421 (2007)].

* Email: yyzhu@nju.edu.cn

# I. INTRODUCTION

Due to the unique dielectric response of metals, the interaction between the electromagnetic waves and microstructured metals has been a subject of a wealth of theoretical and experimental work. For example, by using the periodic arrays of conducting wires and split-ring resonators, left-handed metamaterials have been constructed [1]. In recent years, resonant excitation of surface plasmon polariton (SPP) has sparked much attention [2]. Light that usually travels in a dielectric or vacuum is, due to coupling between light and surface charges, able to propagate along the metal surface with the amplitudes decaying into both sides. This unique feature has generated the SPP-based photonics or plasmonics. Plasmonics possessing both the capacity of photonics and miniaturization of electronics enables it to be an outstanding candidate for the future optoelectronic applications [3].

Besides the propagation of light along metal surface, the light propagation normal to the metal film is another interesting issue. When the film is optically thin (~ skin depth), the energy transfer can be enhanced by the SPP mode [4], and super-resolution imaging can be achieved [5]. And when the film thickness is much larger than the skin depth, the film will be completely opaque. However, it was found by Ebbesen et al. [6] that when subwavelength hole arrays are introduced into the metal film, a surprising effect appears. That is, the transmission of light thought to be very low can be larger than unity when normalized to the area of the holes. This unusual transmission effect has generated great interest in the scientific community [7, 8], partly due to its wide range of potential applications, including subwavelength light sources, quantum information processing, and so on.

Recently, much attention has been devoted to the underlying physics of the phenomenon [9]. In such a metal system, surface plasmon polariton (SPP) mode can be produced by grating coupling and leads to strong enhancement of fields at the metal surface. Naturally it is admitted that the SPP plays a crucial role [6]. However, the experimental results such as the position of transmission peaks do not agree with the SPP model [10, 11]. Even in a recent experiment [12], the attribution of absorption maxima to the SPP mode deserves further deliberation. Theoretically the problem can be treated numerically or analytically. But from numerical calculations it is not obvious what mechanism is involved. In addition, the numerical results deviate significantly from the experiments [13, 14]. And previous analytical results bear a poor prediction of peak width as well as an obscure SPP interpretation [15]. Therefore, the SPP mechanism has received a drastic debate [16-18].

In contrary to SPP model, the composite diffracted evanescent wave (CDEW) model [18] proposed that the transmission anomalies originate just from a unique interference effect at the metal surface (the role of SPP is negligible). That is, constructive interference of evanescent waves at the hole opening gives rise to the transmission maximum and destructive interference the transmission minimum. However, this model cannot explain the dependence of transmission maximum and minimum on the hole size, where the peak position varies with the hole side while the dip position does not [13, 14]. Moreover, their calculation results rely on some fitting parameters, and many factors have not been taken into account. Therefore, the CDEW model cannot give a complete description of transmission properties, implying that the underlying physics remains an open question rather than having been solved.

In this paper, the light transmission through perforated metal film has been analytically studied, and the underlying physics has been discussed. The paper is structured as follows. In section II, an analytical calculation of the zero-order transmission spectrum is presented. In section III, the results of analytical calculations are compared with the experiments, and good agreement has been achieved. Section IV gives a discussion of the underlying physics of the phenomenon. And a short summary is provided in section V.

## II. ANALYTICAL CALCULATIONS

To look into the underlying physics of the transmission, the following case will be considered. Here the metal film with thickness $h$ ($h$ is much larger than the skin depth) is sandwiched between medium 1 and 3, where the permittivity of metal and dielectrics are $\varepsilon_m$, $\varepsilon_1$ and $\varepsilon_3$, respectively. For analytical convenience, we assume that rectangular holes with sides $a \times b$ ($a$ in x-direction and $b$ in z-direction) are cut into the film in a square array of lattice constant $d$. And a TM polarized light (the magnetic field is along the z-axis) is incident upon the metal film and transmits from medium 1 to 3, where x-y plane is the incident plane and $\varphi$ the incident angle. Fig. 1 shows a schematic view of the structure under study.

It is a common knowledge that diffraction modes can be generated when the incident light impinges on the structured metal surface. These diffraction modes will be of great importance in the light transmission. In addition, for the structured metal film, the surface charges can be excited and couple strongly to the electromagnetic fields. These surface plasmons, whether they play positive or negative role, will participate in shaping and simultaneously be marked in the transmission spectrum. Previously, the subwavelength hole walls have been taken as perfect conductors [15, 19, 20]. But the fact is that, in the transmission the electromagnetic waves are strongly confined in the subwavelength holes, which act as the unique channel for the photon tunneling. Then besides the SPP mode formed on the film surface, strong coupling between light and electrons on the hole walls can also be induced. This is the analytical model we used below which, we believe, captures the essence of physics of the enhanced light transmission.

### A. Cavity surface plasmons

As is known, in the optical regime, the metal is not perfect conducting. Then the case is much complex for the rectangular holes with metallic hole walls, where the cavity mode is neither TM nor TE mode. Nonetheless, considering of the polarization of incident light, the dominant in-plane (x-z plane) components of the cavity mode will be $H_z^h$ and $E_x^h$. And correspondingly, the components $H_x^h$ and $E_z^h$ are much smaller and thus can be neglected (this can be well satisfied when $a >> \lambda/\pi\sqrt{-\varepsilon_m} \approx 45\,nm$ ). In the single-mode approximation, by using the surface-impedance boundary condition (SIBC) imposed on the hole walls, the cavity mode can be expressed as

$$H_z^h = (A_0 e^{-iq_0 y} + B_0 e^{iq_0 y})\cos(\alpha x)\cos(\beta z). \tag{1}$$
$$(-a/2 \le x \le a/2,\ -b/2 \le z \le b/2)$$

Here, $A_0$ and $B_0$ are the unknown amplitudes of the downward and upward waves in the nanoholes; $q_0 = \sqrt{k_h^2 - \alpha^2 - \beta^2}$ is the propagation constant of cavity mode, where $k_h = k_0\sqrt{\varepsilon_h}$ ($\varepsilon_h$ is the permittivity of any material filled in the cavities); $\alpha$ and $\beta$ are determined respectively by

$$tg\frac{\alpha a}{2} = \frac{k_0 \varepsilon_h}{i\alpha\sqrt{\varepsilon_m}},\quad tg\frac{\beta b}{2} = \frac{k_0\sqrt{\varepsilon_m}}{i\beta}. \tag{2}$$

When the hole walls are perfect conducting ($\varepsilon_m \to -\infty$), equations (2) lead to $\alpha = 0$ and $\beta = \pi/b$. Correspondingly, the cavity mode is just simplified to the $TE_{01}$ mode. But when the real metal is considered, it is found that $\alpha$ becomes a purely imaginary number ($\alpha \approx i\sqrt{2k_0 \varepsilon_h /(-\varepsilon_m)^{1/2} a}$), and $\beta$ is real but smaller than $\pi/b$ (if the absorption of metal is neglected). That means, the fields in the cavities are propagating in the z direction but evanescent in the x direction; i.e., they are bounded to the hole walls ($x = \pm a/2$). This surface mode confined in the nanoholes, originating from coupling of light to collective oscillation of electrons, can be termed the cavity surface plasmons (CSP). In addition, the propagation constant $q_0$ is also imaginary when the wavelength is much larger than the hole size. Comparing with the tunneling effect of electrons where the electron energy is lower than the potential and the wavevector of de Broglie wave is imaginary, here the photon tunneling effect in the nanoholes can be expected. Due to the presence of CSP mode, $|q_0|$ becomes smaller and thus the photons are easier to pass through the subwavelength holes than that in a perfect-metal case. This efficient channel has been proposed by Popov et al. [21], but the origin was not clarified. Now, it is clear that this CSP-assisted and enhanced photon tunneling effect will contribute actively to the light transmission.

### B. Zero-order transmission

With the use of electromagnetic boundary conditions, the distribution of wave fields in the space has been deduced analytically, which is found to be strongly dependent on a function $F(\lambda)$, where

$$F(\lambda)^{-1} = (1+\theta_1^-)(1+\theta_3^-) \\ - (1-\theta_1^+)(1-\theta_3^+) e^{2iq_0 h}. \tag{3}$$

Here $\theta_j^\pm$ ($j = 1, 3$) is given by

$$\theta_j^\pm = \frac{\sigma \pm \varepsilon_m^{-1/2}}{\sin c(\alpha a/2)} \sum_{n=-\infty}^{+\infty} \frac{w\varepsilon_j g_n s_n}{(\varepsilon_j - \gamma_n^2)^{1/2} + \varepsilon_j \varepsilon_m^{-1/2}}, \tag{4}$$

where, $\sigma = k_0 q_0^{-1}(1-\alpha^2 k_h^{-2})$ is associated with the CSP mode, $w = ab/d^2$ is the duty cycle of the subwavelength holes; the summation on $n$ ($n = 0, \pm 1, \pm 2, ...$) involves all the diffraction orders above or below the metal film, where $g_n = \sin c(k_0 \gamma_n a/2)$, $s_n = (1/a)\int_{-a/2}^{a/2} e^{-ik_0 \gamma_n x} \cos(\alpha x) dx$, $\gamma_n = \sqrt{\varepsilon_1} \sin\varphi + G_n/k_0$, and $G_n = 2\pi n/d$ is the reciprocal lattice vector.

With the function $F(\lambda)$, the zero-order transmission can be written as

$$t_0 = \sqrt{\varepsilon_1 \varepsilon_3} \left| \tau F(\lambda) \right|^2. \tag{5}$$

Here, $\tau$ is a coefficient related to the incident angle, which can be simplified at normal incidence to

$$\tau = \frac{4w\sigma e^{iq_0 h}}{(1+\varepsilon_1^{1/2}\varepsilon_m^{-1/2})(1+\varepsilon_3^{1/2}\varepsilon_m^{-1/2})}. \tag{6}$$

In the following, equations (5) and (6) will be employed to calculate the zero-order transmission efficiency of light.

### III. COMPARING TO EXPERIMENTS

#### A. Comparing to our experiments

In our experiments, optically thick Au or Ag film is coated by sputtering on the cross section of a single-mode optical fiber (NA=0.1). Square holes arranged in a square array ($14\mu m \times 14\mu m$) are fabricated in the film with the focused-ion-beam system (strata FIB 201, FEI company, 30 keV Ga ions). In the measurement, the incident light from an incoherent light source is coupled into the single-mode fiber for which the light can be considered to be incident normal to the metal film. And the zero-order transmission spectrum is obtained with an optical-spectrum analyzer (ANDO AQ-6315A). The measured transmission spectra have been compared with the analytical calculations, and a good agreement has been achieved. In the calculations, the refractive index for air and fiber is 1.0 and 1.46 respectively. And a frequency-dependent permittivity for the metal has been used ($\varepsilon_m = \varepsilon_R + i\varepsilon_I$), where

$$\varepsilon_R = a_1(1-e^{-a_2\lambda})^{a_3}, \quad \varepsilon_I = b_1 + b_2\lambda + b_3\rho^\lambda. \tag{7}$$

In the considered wavelength region ($0.5 \leq \lambda \leq 1.6 \, \mu m$), the associated coefficients fitted with the experimental data [22] are showed in Table 1.

Here, for example, the results for two samples perforated with square holes will be presented. The first sample was made in a Au film, with the lattice constant $d$=761nm, film thickness $h$=220nm, and hole side $a=b$=383nm. And the second sample was made in a much thicker Ag film, where $d$=600nm, $h$=420nm, and $a=b$=270nm. The measured (the circles) and calculated (the line) transmission spectra for the two samples are showed in Fig. 2(a) and 2(b), respectively (Wood's anomaly and SPP resonance have been marked respectively by the thinner and thicker arrows). One can see that the experimental results, including the spectrum shape, the positions of transmission minima and maxima are well reproduced by the calculations. For the first sample, for example, two measured transmission dips locate at the wavelengths 781 and 1129nm, which is very close to the calculated 779 and 1128nm. And the transmission efficiency is greatly enhanced with the peaks positioned at 905 and 1237nm, which agrees respectively with the theoretical values 903 and 1257nm. Moreover, good agreement can also be extended to the width of transmission peaks (or dips), as can be seen from the spectra. For the other sample, similar results can be found (note that some deviations are resulted in the shorter wavelength region, where the single-mode approximation employed in the theory will be less well satisfied). In previous work [15], the analytically calculated peak width is less than 20% of the measured values. Even in a numerical simulation which is found in the literature to be most successful [13, 14], the deviation of peak position (and peak width) between theory and experiment is still up to about 80nm (and 40%).

### B. Comparing to those in literature

To demonstrate the validity of the analytical calculations, the calculated results have been compared with the experiments in literature. And good agreement has been achieved again.

In Ref. [14], Molen et al. studied the zero-order transmission spectrum of rectangular holes (Au film on BK7 glass substrate), where the lattice constant is fixed and hole size is changed. They found that the transmission peak locating at the longer wavelength depends strongly on the aspect ratio. Here our calculated results are displayed in Fig. 3(a) (the structure parameters are chosen from Fig. 1(a) of Ref. [14]). When the aspect ratio varies from 1.5 to 2.0, 2.5, and 3.0, the longer transmission peak shifts from 765 to 813, 870 and 930nm correspondingly; this agrees extremely well with their experiments (763, 828, 881 and 933nm). And the peak width and transmissivity increase with the aspect ratio, as is shown in experiments. And in Ref. [23], Degiron et al. measured the zero-order transmission spectrum of rectangular holes (free-standing Ag film), where the hole size is fixed and the lattice constant is varied instead. The experiments suggest the evolution of transmission spectrum with the change of lattice constant. We have calculated the spectra of three samples, where the film thickness is 400nm, the hole opening is 200nm*260nm, and the period is 400, 450, and 500nm respectively. The results for three samples are showed in Fig. 3(b), which corresponds respectively to the experimental spectrum in Fig. 5 (dotted line), Fig. 3 (solid line), and Fig. 6 (dotted line) of Ref. [23]. From the spectra, one can see that red-shift of transmission peak, peak positions and peak widths agree well with the experiments.

Besides the transmission spectrum, the reflection and absorption spectra of the perorated metal film can provide us with additional insight into the phenomenon. Here we have calculated the spectra with the analytical results, where 200nm thick Au film (glass substrate) and normal incidence are assumed, the lattice constant is 700nm, and the hole side of square holes is 300nm. Fig. 4(a) and 4(b) shows the calculated transmission, reflection and absorption spectra for light incident from air and glass side, respectively. The results suggest that (in the zero-order regime) the transmission maxima are associated with reflection minima and absorption maxima, and that transmission minima are associated with reflection maxima and absorption minima. These characters are just the same as that of the experiments of Barnes et al. [12]. In addition, Fig. 4 suggests that the transmission spectrum is identical whether air or glass side is illuminated, which is in accordance with the observations. Our results also show that it is nonreciprocal for the reflection and absorption spectra for the two cases; this point has been suggested by the experiments [12, 24]. Overall, our analytical result agrees well with the experiments, and thus reveals the underlying physics of the light transmission.

## IV. DISCUSSIONS

Equation (5) suggests that the light transmission is governed by the function $F(\lambda)$, which is related to both the CSP mode and the diffraction orders. And generally the positions of transmission dips and peaks correspond respectively to the minima and maxima of $|F(\lambda)|$. When the denominator in equation (4) approaches the zero, i.e., $(\varepsilon_j - \gamma_n^2)^{1/2} + \varepsilon_j \varepsilon_m^{-1/2} = 0$, the transmission dip can be reached ($F(\lambda) \to 0$). The equality can be rewritten in a familiar form:

$$k_0 \sqrt{\varepsilon_1} \sin\varphi + G_n = \pm k_{spp}, \qquad (8)$$

with

$$k_{spp} = \frac{\omega}{c} \sqrt{\frac{\varepsilon_j(\varepsilon_m^2 - \varepsilon_j^2)}{\varepsilon_m(\varepsilon_m + \varepsilon_j)}} \approx \frac{\omega}{c} \sqrt{\frac{\varepsilon_m \varepsilon_j}{\varepsilon_m + \varepsilon_j}}. \qquad (9)$$

Where $\varepsilon_j^2$ is much smaller than $\varepsilon_m^2$. This is just the condition for SPP resonance on the surface of the metal film (the approximated SPP dispersion is caused by the SIBC imposed on the film surface). Therefore, the SPP mode is responsible for the transmission dips (which can be termed the SPP minima); this is identical to the experiment (see the thicker arrows in Fig. 2).

We notice in a recent calculation that the dispersion of SPP on a perforated metal film has been proposed to be strongly modified [25]. There the so-called dispersion curve obtained by searching for the poles of scattering matrix is linked actually to all the diffraction orders as well as cavity modes. Thus it cannot be attributed merely to the SPP wave. Moreover, the change of dispersion by more than 10% [25] is difficult to understand, considering the fact that the SPP dispersion on slit arrays is within 1% of the flat case [18]. Here the theory and experiment show that when the resonance condition (equation (8)) is satisfied a sudden decrease in the transmission will be induced. The results strongly suggest that the SPP mode (the inherent surface resonance) of a flat metal surface still survives and acts on the perforated metal film, i.e., the residual metal surface.

The negative role of SPP mode has been suggested in the subwavelength slits [17, 26] and admitted by Ebbesen et al. [6]. This point hold even for two-dimensional holes is not difficult to understand. When the resonance condition on incident side is satisfied, some evanescent diffraction mode will couple strongly to the SPP wave and others are suppressed, where the photons seem to be absorbed on the surface. Nevertheless, due to the coupling between light and electrons, the enhancement of fields associated with SPP resonance is limited to the metallic boundary (the hole opening is not favored). Moreover, the average energy flow of SPP wave travels parallel to the metal film (which is much thicker than the skin depth). Hence, the SPP mode does not give rise to any efficient transmission of light. Additionally, when the SPP resonance on the other side is excited, the remaining diffraction modes (exit side) will be suppressed accordingly, including the zero-order transmission.

An important result predicted by the theory is that the Wood's anomaly does not give rise to transmission minimum, which is contrary to the conventional idea [10]. Since $\left|\varepsilon_j \varepsilon_m^{-1/2}\right| \ll 1$, when $\varepsilon_j - \gamma_n^2 = 0$ (this is the condition for Wood's anomaly), the transmission is low but not zero (see the thinner arrows in Fig. 2). Physically, in the case of Wood's anomaly the related diffraction order becomes tangent to the metal surface and unable to transmit. But the remaining diffraction modes have not been suppressed completely, which couple with the CSP mode and result in a weak transmittance. In fact, the position of Wood's anomaly is always slightly smaller than that of transmission dip. For the first-order reciprocal vector and normal incidence, they position respectively at $\lambda_W = \sqrt{\varepsilon_j}\, d$ and $\lambda_S = \sqrt{\varepsilon_m \varepsilon_j /(\varepsilon_m + \varepsilon_j)}\, d$. Only when the metal film is perfect conducting, can they be identical.

In addition, the analytical results show that the amplitude of CSP mode ($A_0$ and $B_0$ in equation (1)) is proportional to $F(\lambda)$. Thus the CSP mode is greatly suppressed when the SPP wave is resonantly excited ($F(\lambda) \to 0$). However, when the wavelength is far from the SPP resonance, a coupling between the evanescent diffraction modes and CSP mode will be established. The evanescent diffraction modes can propagate along the whole surface, including the nonmetallic interface at the hole opening, where coupling occurs through the matching of the fields. Then the diffraction modes on the incident side will emit the photons that tunnel via the CSP mode to the other side, which further couples to the outgoing light. This coupling can be strongly resonant when $|F(\lambda)|$ reaches its maximum. Beyond the cutoff wavelength of CSP mode which is interested here, the second term on the right-hand side of Eq. (3) can be neglected. Then the condition for resonance can be simplified to $1 + \theta_j^- = 0$, which requires both $q_0$ and $\sqrt{\varepsilon_j - \gamma_n^2}$ being imaginary numbers. When $q_0$ is imaginary, the CSP mode is evanescent and bounded to the hole opening. And when $\sqrt{\varepsilon_j - \gamma_n^2}$ is imaginary, the diffraction mode is evanescent and bounded to the film surface. Therefore, they couple strongly to each other, leading to great enhancement of CSP mode and evanescent diffraction modes on either side. Correspondingly, a strong power flow normal to the metal film is supported ($S_h \propto |F(\lambda)|^2$) and the energy is transferred efficiently from one side to the other side. This differs significantly from the case when the SPP is resonantly excited. Just because of the dependence of transmission peaks on CSP and diffraction orders, a variation of hole size (relating to CSP mode) or lattice constant (relating to diffraction modes) may shift the position of peaks, as demonstrated by the experiments [14, 23].

To help understand the above results, we have calculated the amplitudes of diffraction modes with the parameters described in Fig. 4 (the magnetic field of incident light is set as unity). The results are displayed in Fig. 5 (a) and 5 (b), which corresponds respectively to the first- and second-order diffraction modes (higher orders are not present here). One can see that the fields are strong at both transmission minima and maxima. But there are some differences between them. At the transmission minima (indicated by the arrows), the SPP resonance is excited and only one diffraction order is very strong. And at the transmission maxima (which is far from SPP resonance), instead, all the diffraction orders on both sides are enhanced with the high-orders being much weak. On the other hand, the diffraction modes appear to be much stronger at the transmission maxima than that at the minima. With these features, it is not difficult to understand why the transmission maxima are associated with the absorption maxima [12].

The optical transmission through a perfect metal film perforated with square holes has also been calculated (the line in Fig. 6a), where the SPP and CSP modes are both absent. One can see that the transmission dips and peaks are still supported, implying that the surface plasmons are not necessary for transmission. However, the case is indeed different when they are present (the circles in Fig. 6a). On the one hand, the dip position is shifted from Wood's anomaly to SPP resonance (the SPP contribution). On the other hand, the position of peaks is red-shift and simultaneously the peak width is greatly enlarged (the CSP contribution). It is also found that when the metal film becomes thicker, the transmission efficiency for a real metal (the circles in Fig. 6b) is even higher than that of a perfect one (the line in Fig. 6b), which confirms the active role of CSP mode.

## V. CONCLUSIONS

The light transmission through perforated metal films has been studied theoretically, considering of both diffraction modes (which is inherent in periodic structures) and surface plasmons (surface plasmons on film surface as well as hole inside, which is inherent in metallic materials). A good agreement between analytical calculations and experiments has been achieved. Although the results presented here are based for convenience on square and rectangular holes, the underlying physics is general. In the light transmission, the diffraction modes interact with the surface plasmons: strong coupling of diffraction modes to SPP results in the transmission minima and to CSP the transmission maxima. In this framework, they have been treated fairly: the diffraction modes are in dominant and driving place, while the SPP and CSP mode are in driven place. Correspondingly, a dual role is played by the surface plasmons.

## ACKNOWLEDGEMENTS

This work was supported by the State Key Program for Basic Research of China (Grant No. 2004CB619003), by the National Natural Science Foundation of China (NNSFC, Grant Nos. 10523001 and 10474042). C.P. Huang would also like to acknowledge partial support from the NNSFC under Grant No. 60606020.

**References**

[1] D.R. Smith, W.J. Padilla, D.C. Vier, S.C. Nemat-Nasser, and S. Schultz, Phys.Rev.Lett. **84**, 4184 (2000).

[2] W.L. Barnes, A. Dereux, and T.W. Ebbesen, Nature **424**, 824 (2003).

[3] E. Ozbay, Science **311**, 189 (2006).

[4] P. Andrew and W.L. Barnes, Science **306**, 1002 (2004).

[5] N. Fang, H. Lee, C. Sun, and X. Zhang, Science **308**, 534 (2005).

[6] T.W. Ebbesen, H.J. Lezec, H.F. Ghaemi, T. Thio, and P.A. Wolff, Nature **391**, 667 (1998).

[7] J.R. Sambles, Nature **391**, 641 (1998).

[8] S. Fasel, F. Robin, E. Moreno, D. Erni, N. Gisin, and H. Zbinden, Phys.Rev.Lett. **94**, 110501 (2005).

[9] T.D. Visser, Nature Phys. **2**, 509 (2006).

[10] H.F. Ghaemi, T. Thio, D.E. Grupp, T.W. Ebbesen, and H.J. Lezec, Phys.Rev.B **58**, 6779 (1998).

[11] J. Bravo-Abad, A. Degiron, F. Przybilla, C. Genet, F.J. García-Vidal, L. Martín-Moreno, and T.W. Ebbesen, Nature Phys. **2**, 120 (2006).

[12] W.L. Barnes, W.A. Murray, J. Dintinger, E. Devaux, and T.W. Ebbesen, Phys.Rev.Lett. **92**, 107401 (2004).

[13] K.J. Klein Koerkamp, S. Enoch, F.B. Segerink, N.F. van Hulst, and L. Kuipers, Phys.Rev.Lett. **92**, 183901 (2004).

[14] K.L. Van der Molen, K.J. Klein Koerkamp, S. Enoch, F.B. Segerink, N.F. van Hulst, and L. Kuipers, Phys.Rev.B **72**, 045421 (2005).

[15] L. Martin-Moreno, F.J. Garcia-Vidal, H.J. Lezec, K.M. Pellerin, T. Thio, J.B. Pendry, T.W. Ebbesen, Phys.Rev.Lett. **86**, 1114 (2001).

[16] M.M.J. Treacy, Phys.Rev.B **66**, 195105 (2002).

[17] Q. Cao and P. Lalanne, Phys.Rev.Lett. **88**, 057403 (2002).

[18] H.J. Lezec and T. Thio, Opt. Express **12**, 3629 (2004).

[19] J.B. Pendry, L. Martin-Moreno, and F.J. Garcia-Vidal, Science **305**, 847 (2004).

[20] F.J. Garcia-Vidal, E. Moreno, J.A. Porto, and L. Martin-Moreno, Phys.Rev.Lett. **95**, 103901 (2005).

[21] E. Popov, M. Neviere, S. Enoch, and R. Reinisch, Phys.Rev.B **62**, 16100 (2000).

[22] E.D. Palik, Handbook of optical constants in solids (Academic, Boston, 1991).

[23] A. Degiron and T.W. Ebbesen, J.Opt.A Pure Appl. Opt **7**, S90 (2005).

[24] M. Sarrazin and J.P. Vigneron, Phys.Rev.B **70**, 193409 (2004).

[25] P. Lalanne, J.C. Rodier, and J.P. Hugonin, J.Opt.A: Pure Appl.Opt. **7**, 422 (2005).

[26] H. Lochbihler, Phys.Rev.B **50**, 4795 (1994).

**Figure captions**

Fig.1 Schematic view of the subwavelength holes in a metal film, where the film thickness is $h$, and rectangular holes ($a \times b$) are arranged in a square array ($d \times d$). The light lying in x-y plane is incident with a tilt angle $\varphi$ upon the metal surface, and transmits from medium 1 to 3 (the magnetic polarization of light is in the z direction).

Fig.2 Measured (the open circles) and calculated (the solid line) zero-order transmission spectra of the metal film perforated with square holes. (a) Au: d=761nm, a=b=383nm, and h=220nm; (b) Ag: d=600nm, a=b=270nm, and h=420nm. The thicker and thinner arrows indicate the positions of SPP resonance and Wood's anomaly, respectively.

Fig.3 Calculated zero-order transmission spectra of rectangular holes. (a) Au (BK7 substrate): d=425nm, h=200nm, and the hole size is varied with the aspect ratio being 1.5, 2.0, 2.5, 3.0 respectively (the hole area is kept at 22500nm$^2$, see Ref. [14]); (b) Ag (free-standing): h=400nm, a=200nm, b=260nm, and the lattice constant is varied with d= 400, 450, 500nm respectively (see Ref. [23]).

Fig.4 Calculated zero-order reflectance, transmittance and absorbance of the perforated Au film: (a) incident from the air side, and (b) incident from the glass side. Here, d=700nm, a=b=300nm, and h=200nm.

Fig.5 Amplitudes of diffraction modes: first-order (a) and second-order (b) reflected ($|R_{1,2}|$) and transmitted ($|T_{1,2}|$) waves (the circles represent the zero-order transmission spectra). Here d=700nm, a=b=300nm, h=200nm (glass substrate), and the light is incident from the air side.

Fig.6 Zero-order transmission spectra of the real as well as perfect Au film: (a) h=200nm, and (b) h=300nm. Here the thick lines are associated with the real metals, and the thin lines are with the perfect conductors. In all cases, d=700nm, and a=b=300nm (glass substrate).

Table 1. The fitting parameters for permittivity of metals

|       | Au        | Ag         |
|-------|-----------|------------|
| $a_1$ | -128.4231 | -1303.6078 |
| $a_2$ | 2.2138    | 0.3243     |
| $a_3$ | 8.6838    | 2.6134     |
| $b_1$ | -11.4226  | -2878.5803 |
| $b_2$ | 14.3676   | 260.7128   |
| $b_3$ | 90.3912   | 2884.3145  |
| $\rho$ | 0.0062   | 0.9086     |

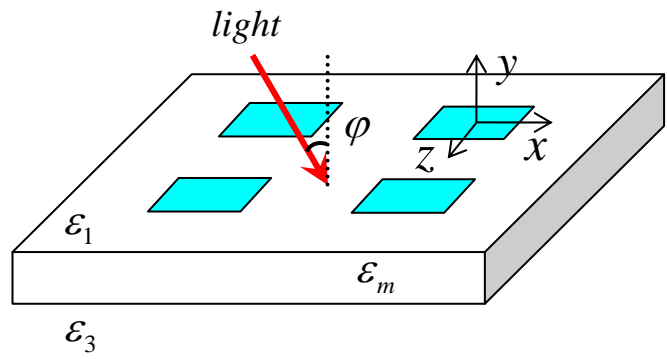

**Fig. 1**

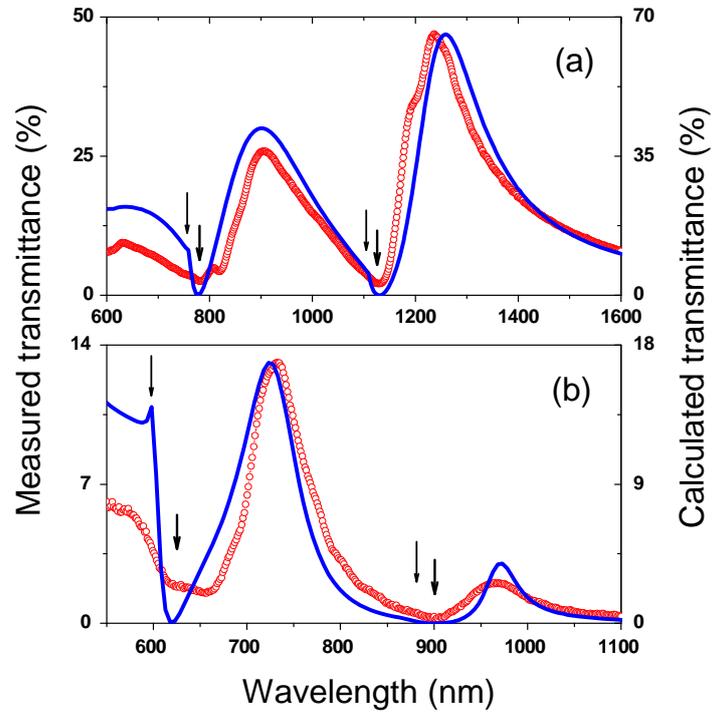

**Fig. 2**

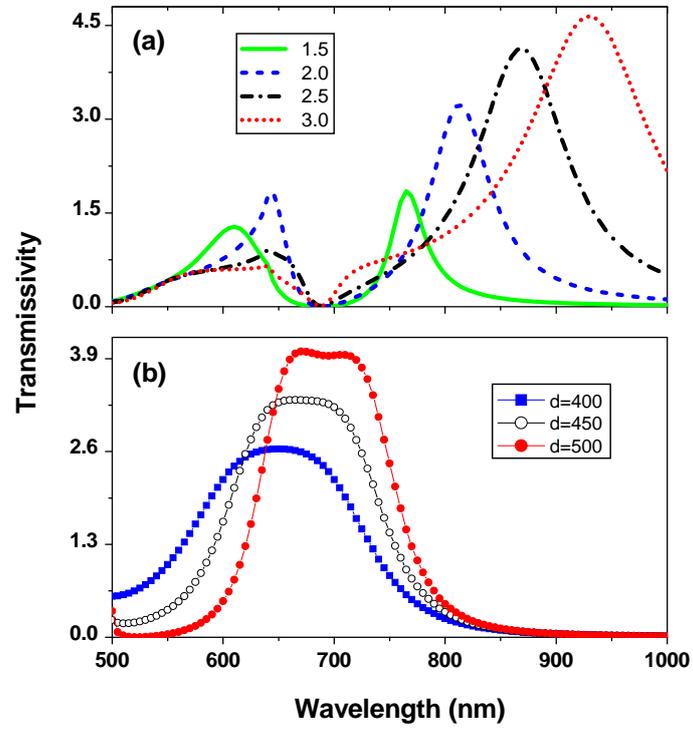

**Fig. 3**

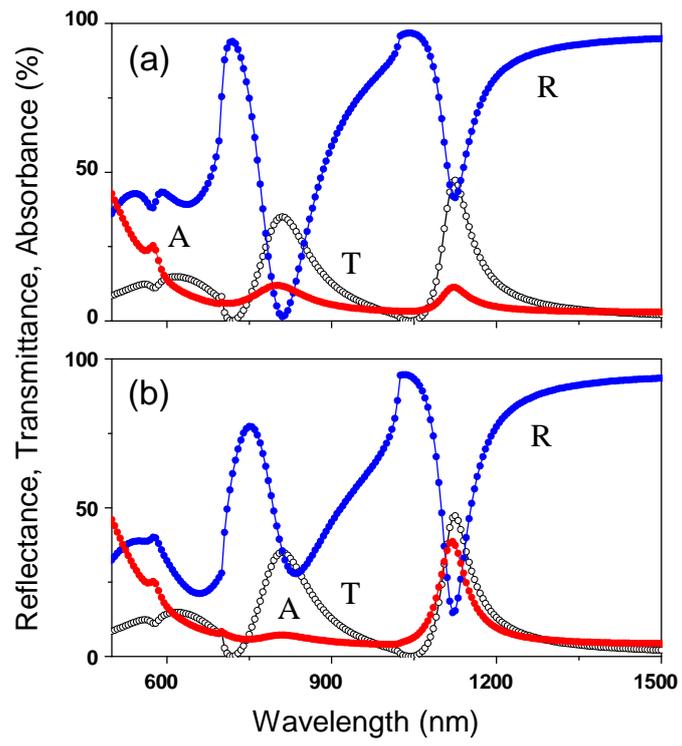

**Fig. 4**

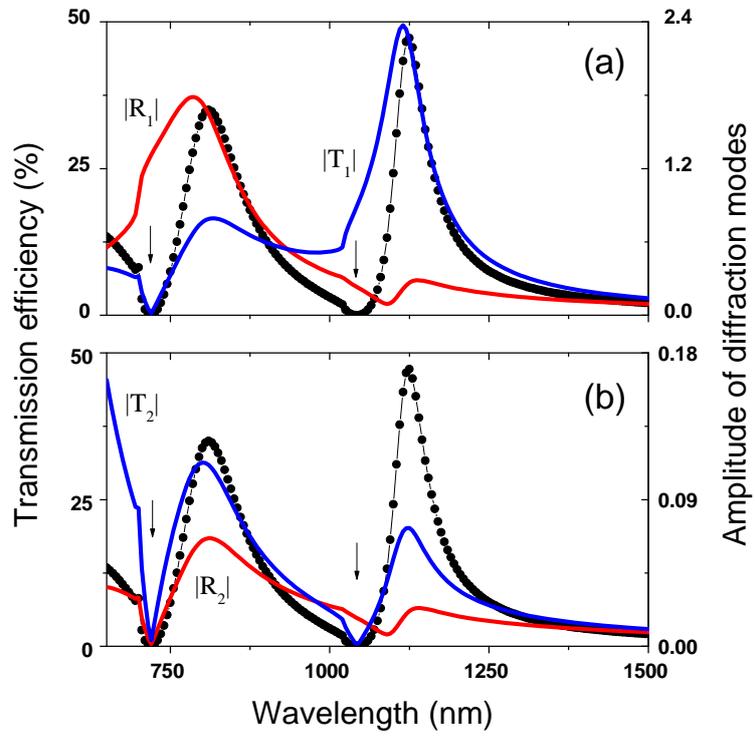

**Fig. 5**

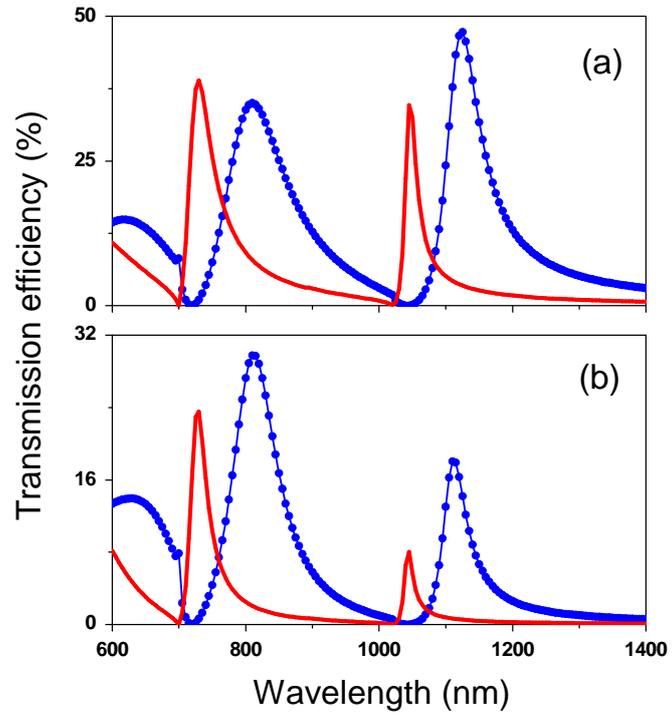

**Fig. 6**